\documentclass[twocolumn,prl,preprintnumbers,showpacs,amsmath,amssymb]{revtex4}
\usepackage{graphicx}
\usepackage{dcolumn}
\usepackage{txfonts}
\usepackage{xcolor}
\usepackage{amssymb}
\usepackage{amsmath}
\usepackage{latexsym}
\usepackage{epsfig}
\usepackage{amsbsy}
\usepackage{array}
\usepackage{tabularx}

\begin{document}

\title []{Topological Metal of NaBi
with Ultralow Lattice Thermal Conductivity and Electron-phonon
Superconductivity}

\author{Xing-Qiu Chen}
\email[Corresponding author:]{xingqiu.chen@imr.ac.cn}
\author{Ronghan Li}
\author{Yan Sun}
\author{Xiyue Cheng}
\author{Dianzhong Li}
\author{Yiyi Li}

\affiliation{Shenyang National Laboratory for Materials
Science, Institute of Metal Research, Chinese Academy of Sciences,
Shenyang 110016, China}

\date{\today}

\begin{abstract}
By means of first-principles and \emph{ab initio} tight-binding
calculations, we found that the compound of NaBi is a
three-dimensional non-trivial topological metal. Its topological
feature can be confirmed by the presence of band inversion, the
derived effective Z$_2$ invariant and the non-trivial surface states
with the presence of Dirac cones. Interestingly, our calculations
further demonstrated that NaBi exhibits the uniquely combined
properties between the electron-phonon coupling superconductivity in
nice agreement with recent experimental measurements and the
obviously anisotropic but extremely low thermal conductivity. The
spin-orbit coupling effects greatly affect those properties. NaBi
may provide a rich platform to study the relationship among metal,
topology, superconductivity and thermal conductivity.
\end{abstract}
\pacs{71.20.-b, 71.15.-m, 71.38.-k, 73.20.-r}

\maketitle

Because the topological concept was successfully introduced into
insulators, various insulators can be classified into topological
trivial and non-trivial states\cite{Kane1, Shou-Cheng Zhang1,
Shou-Cheng Zhang3,Kane2}, in which topological insulators are
highlighting an exciting scientific frontier of the topological
electronic states. In analog of insulators, semimetals can also be
classified from topological points as trivial semimetals and
topological non-trivial semimetals (TSMs). Among TSMs, there are two
classes of peculiar materials, topological Dirac semimetals
(TDSs)\cite{Dirac1,Wang,Cheng,Wang2} and topological Weyl semimetals
(TWSs)\cite{Weyl1,Weyl2,Weyl3,Weyl4,Weyl5,
Weyl6,Weyl8,Weyl10,Weyl11,Weyl14}, in which Fermi surfaces are
consisted of isolated Fermi points in lattice momentum space. In
general, the TDSs are predicted to exist at the critical phase
transition point from a normal insulator and a topological one
through the spin-orbit coupling effect or by tuning the chemical
composition \cite{Murakami,Young2}. However, such bulk Dirac points
are occasionally degeneracies and not stable. Interestingly, very
recently the systems of the $P$6$_3$/$mmc$-Na$_3$Bi \cite{Wang,
Cheng, Liu, Xu2013} and $\beta$-BiO$_2$ \cite{Young} and
Cd$_3$As$_2$ \cite{Wang2,Neupane,Borisenko,ylchen2,Jeon} have been
predicted theoretically and then Na$_3$Bi and Cd$_3$As$_2$ have been
experimentally confirmed to be robust TDSs protected by crystal
symmetry. TWSs have been theoretically suggested to appear in
skutterudite-structure pnictides\cite{Weyl6}, pyrochlore
iridates\cite{Weyl10}, doped compound
Hg$_{1-x-y}$Cd$_x$Mn$_y$Te\cite{Weyl8} and some constructed
heterostructures\cite{Weyl11}, but to date no experimental
verification has been achieved.

Certainly, there is no doubt that the topological concept can be
also introduced into metals. Hence, metals would be also classified
into two typical types of trivial metals (Ms) and non-trivial
topological metals (TMs). In fact, many studies have been focused on
the realization and the properties of TMs
\cite{TM1-1D-Bi,TM3-half-Heusler,TM4-half-Heusler,
TM7-BiSb,TM9-(Bi1-xInx)2Se3,TM10-TiS2-xTex}. In general, the TMs can
be achieved just by the effects of imperfections (\emph{i.e.},
chemical doping, strain engineering, heterostructure, etc) on
topological insulators. However, to date for this search of native
TMs (without any doping and strain applications) no material in
reality has been reported successfully. It needs to be emphasized
that TMs would indeed extensively exist. Nevertheless, because the
topological non-trivial states of specified TMs' surface crossing
the Fermi level can be easily mixed by trivial metallic bands, the
real realization of TMs indeed poses a challenge.

\begin{figure}[hbt]
\centering
\includegraphics[width=0.44\textwidth]{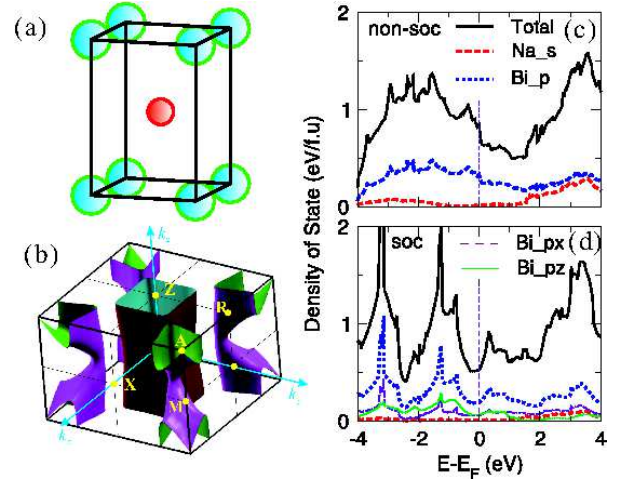}
\caption{Structure, Fermi surface and densities of states
(DOSs) of NaBi.
(a) The unit cell, (b) the Fermi surface with the SOC effect,
(c) and (d) the derived total and projected DOSs without and with
the SOC effect, respectively.}
\label{fig1}
\end{figure}

Within this context, through first-principles calculations with the
framework of Density Functional Theory (DFT) by employing the VASP
code \cite{online,vasp}, here we reported a native 3D TM, NaBi,
which exhibits the combined interesting properties of the
electron-phonon induced superconductivity and the obviously
anisotropic but extremely low bulk thermal conductivity. Its
topological feature has been analyzed according to the band
inversion occurrence between Na-$s$ and Bi-$p$ orbits at the
$\Gamma$ point, the Z$_2$ number based on the derived parities, and
the two selected surface non-trivial helical states. Without (with)
the spin-orbit coupling (SOC) effect the superconducting transition
temperature of $T_c$ is derived to be 1.82-2.59 (2.92-3.75) Kelvin
from the electron-phonon coupling strength $\lambda$ = 0.71 (0.84)
and the average velocity $<\omega>_{ln}$ = 40.8 (38.7) cm$^{-1}$,
agreeing well with the experimental findings \cite{Kushwaha,Joseph}.
In addition, by considering phonon vibrational eigenvalues in the
whole of Brillioun zone (BZ) and the phonon relaxation time derived
from third-order force constants, we have further revealed that NaBi
exhibits an extremely low lattice thermal conductivity but an
obviously anisotropic feature of $\kappa_\omega^{a-axis}$ = 3.98 $W
m^{-1} K^{-1}$ along the \emph{a}-axis and $\kappa_\omega^{c-axis}$
= 1.53 $W m^{-1} K^{-1}$ along the \emph{c}-axis at room
temperature, respectively.

As early as 1932, the compound of NaBi was synthesized to
crystallize in a body-centered tetragonal CuAu-type structure (the
space group of \emph{P}4/\emph{mmm}, No.123, see Fig. \ref{fig1}(a))
with Na at the 1\emph{d} (1/2, 1/2, 1/2) site and Bi at the
1\emph{a} (0, 0, 0) site \cite{NaBi}. The optimized DFT lattice
constants \cite{online} of NaBi at the ground state, $a$ = 3.4116
\AA\, and $c$ = 4.9530 \AA\,, are in nice agreement with the
experimental lattice constants \cite{NaBi} ($a$ = 3.46 \AA\, and $c$
= 4.80 \AA\,). As illustrated in Fig. \ref{fig1}(c and d), NaBi is a
typical metal. The SOC inclusion results in several apparent
features. Without the SOC inclusion, the Fermi level lies in the
declining shoulder of the densities of states (DOS), indicating a
relatively high state of N(E$_F$) = 0.85 states eV$^{-1}$
f.u.$^{-1}$. In contrast, the SOC inclusion significantly reduces
the N(E$_F$) to 0.52 states eV$^{-1}$ f.u.$^{-1}$, due to the fact
that the Fermi level now stays at the valley of the pseudogap. In
addition, from Fig. \ref{fig1}(d) in the occupied states of the DOS
profile the SOC effect even induces the appearance of two obvious
peaks dominated by Bi-$p$-like states at about -3 eV to -1 eV below
the Fermi level, respectively. The presence of those features
indicates the significance of the SOC effect for NaBi.

\begin{figure}[hbt]
\centering
\includegraphics[width=0.48\textwidth]{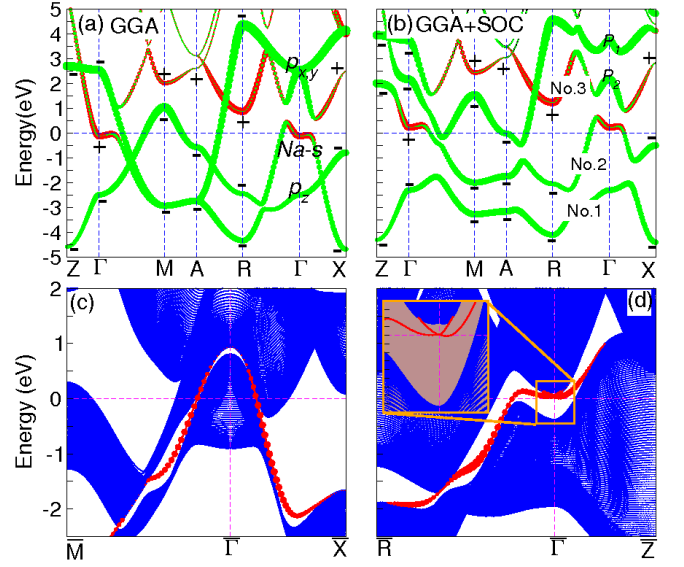}
\caption{Bulk and surface electronic band structures of NaBi. (a)
and (b) the DFT electronic band structures along the high-symmetry
points without and with SOC, respectively. The signs ''+'' and ''-''
denote the parities of bands at the time-reversal invariant momenta
(TRIMs). (c) and (d) corresponds to the (001) and (100) surface
electronic band structures derived from the tight-binding model
(supplementary materials). The red dots show the helical
spin-resolved metallic states on the surfaces. Inset of the panel
(d) displays the surface Dirac cone with the enlarged scale around
the Fermi level at the $\Gamma$ point. } \label{fig2}
\end{figure}

The SOC effect is even more obvious from the electronic band
structures in Fig. \ref{fig2}(a and b). Firstly, without the SOC
inclusion the three bands (as marked by No.1-3 in Fig.
\ref{fig2}(a)) around the Fermi level heavily overlap each other
along some high-symmetry lines. The large SOC effect results in
their separations, as evidenced in Fig. \ref{fig2}(b). It interprets
well as to why two main peaks (corresponding to No.1 and No.2 bands)
occur in the occupied states of the DOS profile (Fig.
\ref{fig1}(d)). In addition, due to the SOC separation between No.2
and No.3 the Fermi level now locates at the valley of the pesudogap.
Secondly, from Fig. \ref{fig2}(a), because of the tetragonal
symmetry with $c$ $>$ $a$ and Bi atoms separated by the
body-centered Na atom, at $\Gamma$ the Bi $p$$_z$ orbital is lower
in energy than both the degenerated Bi $p$$_{x,y}$ orbital and the
Na-$s$ orbital. In particular, the band inversion and the
anti-crossing feature between Na-$s$ and Bi-$p_{x,y}$ orbitals occur
around $\Gamma$, showing a nontrivial gap of about 2.5 eV even
without the SOC effect. It uncovers that this feature is indeed
induced by both the crystal symmetry and the crystal field effect.
Furthermore, under the SOC effect and the D$_{4h}^1$ symmetry the
doubly degenerated Bi-$p$$_{x,y}$ orbitals are further split into
$|$$P_{x,y}^{-}$,$\pm$$\frac{3}{2}$$>$ (as marked by $P_1$ in Fig.
\ref{fig2}(b)) and $|$$P_{x,y}^{-}$,$\pm$$\frac{1}{2}$$>$ (as marked
by $P_2$ in Fig. \ref{fig2}b) states. This leads to a reduced
non-trivial gap of about 1.1 eV between
$|$$P_{z}^{-}$,$\pm$$\frac{1}{2}$$>$ and Na $s$
($|$$s_\frac{1}{2}^{+}$,$\pm$$\frac{1}{2}$$>$) states at $\Gamma$.
Despite of the existence of the non-trivial gap, it is intrinsically
different from topological insulators and topological semimetal
since NaBi is a typical metal. Therefore, it would be extremely
interesting to see whether or not NaBi is a non-trivial TM.

To answer this problem, the most important aspect is to elucidate
whether or not the continuous energy gap exists between No.2 and No.
3 bands in the whole BZ in Fig. \ref{fig2}(b). On the one hand, we
have performed the band structure calculations using a very dense
$k$-mesh set (in total 187836 $k$-point number in the whole BZ, see
Ref.\cite{online}) and the results demonstrated these two bands
never touch each other at any $k$-point and, on the other hand, the
calculations even uncovered that the smallest energy gap between
No.2 and No.3 bands is about 0.08 eV at the four equivalent
($\pm$0.3461$\times$$\frac{2\pi}{a}$,
$\pm$0.3494$\times$$\frac{2\pi}{a}$, 0.5$\times$$\frac{2\pi}{c}$)
points in the \emph{k}-space $k_z$ = 0.5$\times$$\frac{2\pi}{c}$
plane (here, $a$ and $c$ are the lattice constants)\cite{online}. We
have also constructed the tight-binding (TB) model Hamilton
according to the DFT band structure with the SOC inclusion to
further calculate the Berry phase of each energy band in the $k_z$ =
0.5$\times$$\frac{2\pi}{c}$ plane. The result uncovers that the
Berry phase of the No.2 band is zero, thereby evidencing that the
No.2 band never touches the No.1 and No.3 bands. These results fully
evidence the existence of a continuous gap between No.2 and No.3
bands in the whole BZ. Given the fact that the non-trivial gap
exists between these two bands at $\Gamma$ , we can further derive
the topological invariant, according to the Berry curvature and
connection \cite{Fu2006}. Interestingly, for the center-symmetric
structure (with the inversion symmetry) that NaBi crystallizes in,
the effective Z$_2$ invariant can be obtained in terms of the method
proposed by Fu and Kane \cite{Fu2007}. Because the bands below  No.1
band are fully filled and far away in energy, the topological order
just depends on the No.1 and No.2 bands starting from No.1 band
around the Fermi level. As shown in Fig. \ref{fig2}(a and b), the
product of the parities at the eight time-reversal invariant
momentums (TRIMs) is -1, corresponding to $Z_2$ of (1; 0 0 0). It
indicates that NaBi is a strong 3D TM with the presence of the
topological non-trivial states.

We have further examined the intrinsic surface properties of NaBi.
In principles, in similarity to topological insulators, TMs would
have an odd number of Dirac cones to appear at any surface
orientation because the topological order exists. However, for TMs
the behaviors can be highly complex, mainly because the surface
Dirac cones perhaps submerge into the bulk metallic states.
Therefore, in some orientations it would have no chance to see the
presence of surface Dirac cones for TMs. To prove these
expectations, we shall now compute the band dispersions for the
(001) and (100) surfaces using the {\em ab initio} TB model. The
{\em ab initio} TB model is constructed by downfolding the bulk
energy bands, obtained by first-principles calculations using
maximally localized Wannier functions (MLWFs). The MLWFs are derived
from atomic $p$-like and $s$-like states. The surface slab models
(with the terminations of Bi atoms) for the (001) and (100) surfaces
have been constructed with the thickness of 199 and 399 atomic
layers, respectively. The results of the TB calculations are
summarized in see Fig. \ref{fig2}(c and d). For the (001) surface,
the surface electronic bands (as marked by the solid red circles)
connecting the bulk electronic states derived from the No.2 and No.3
bands in Fig. \ref{fig2}(b) cross the Fermi level only once (odd
number) for both $\overline{M}$-$\overline{\Gamma}$ and
$\overline{\Gamma}$-$\overline{X}$. In addition, for this surface no
Dirac cone appears because the surface electronic bands at
$\overline{\Gamma}$ mix totally with the bulk electronic bands
stemmed from the No. 3 band. However, the different behavior has
been observed for the (100) surface [see Fig. \ref{fig2}b]. At
$\overline{\Gamma}$ the clear Dirac cone appears with surface
non-trivial states (as marked by solid red circles) which only once
cut the Fermi level in the $\overline{R}$-$\overline{\Gamma}$
direction. In the $\overline{\Gamma}$-$\overline{Z}$ direction,
there is no crossing at the Fermi level because the surface
non-trivial states in this direction submerges into the bulk band
states derived from the No.2 band. From the viewpoint of the
topology, the cutting number of the Fermi level can be adjusted in
different odd number just by shifting Fermi energy (such as chemical
electronic and hole doping treatments). All these facts further
evidence that NaBi is a 3D non-trivial TM.

\begin{figure}[hbt]
\centering
\includegraphics[width=0.45\textwidth]{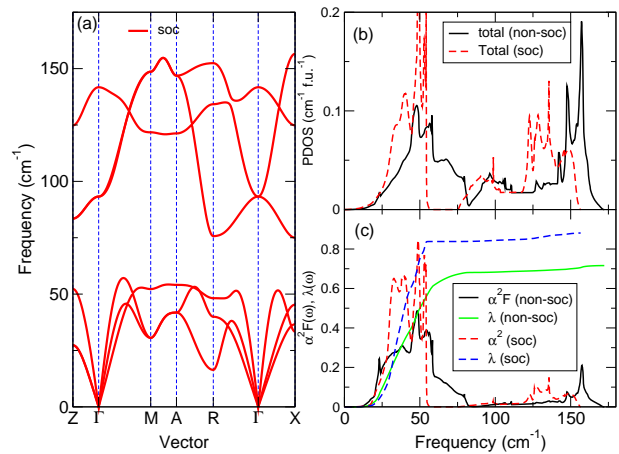}
\caption{Phonon dispersion and electron-phonon coupling strength of
NaBi with the SOC inclusion. (a) Phonon dispersion curves along the high
symmetry lines of the BZ with the SOC inclusion, (b) total and projected
phonon density of states (PHDOS) in NaBi with/without the SOC
inclusion, (c) Eliashberg function and the strength of the
electron-phonon coupling with/without the SOC inclusion.}
\label{fig3}
\end{figure}

We have utilized the linear response theory and fine $k$ and $q$
meshes \cite{online,qe} to calculate the phonon dispersion, phonon
density of states (PHDOS), Eliashberg function
($\alpha^2$F($\omega$)), and the strength of the electron-phonon
($e$-ph) coupling ($\lambda$($\omega$)) with and without the SOC
inclusion. The phonon spectrum and the phonon densities of states in
Fig. \ref{fig3}(a and b) can be divided into two main regions with
mostly Bi (but also slightly mixed with Na) modes (0 - 60 cm$^{-1}$
for SOC and 0 - 85 cm$^{-1}$ for non-SOC) and highly pure Na modes
(80 -- 155 cm$^{-1}$) for SOC and 85 -- 170 cm$^{-1}$ for non-SOC).
As can be inferred for the phonon DOSs in Fig. \ref{fig3}(b), the
SOC inclusion results in the average softening of over 15\% for the
transverse modes, and about 10\% for the longitudinal ones. The
Eliashberg function integrates to a large $e$-ph coupling strength
$\lambda$ = 0.72 (0.84) without (with) the SOC inclusion but gives
the highly low logarithmic average $<\omega>_{ln}$ = 40.9 (38.7 for
SOC) cm$^{-1}$. Although $\lambda$ is very close to the value of
$\lambda$$\sim$0.8 for MgB$_2$ which mainly comes from
high-frequency boron modes \cite{MgB2}, from Fig. \ref{fig3}(c) it
is very clear that nearly over 95\% of $\lambda$ in NaBi is
generated by the dominated Bi modes in the low-frequency acoustic
branches. Strikingly, $<\omega>_{ln}$ in NaBi is found to be only
one tenth of the MgB$_2$ value of $\sim$ 450 cm$^{-1}$ \cite{MgB2}.
Using the Allen-Dynes formula \cite{ADF} and typical $\mu$ of
0.14-0.10 we further estimate the $T_c$ in NaBi to be 1.82-2.59 K
(2.92-3.75 K for SOC) (see Table \ref{tab1}). Although the estimated
data without the SOC inclusion yields a perfect agreement with the
experimental data \cite{Kushwaha}, the SOC inclusion indeed exhibits
a significant effect on these superconducting parameters.

In particular, it needs to be emphasized that the compound of NaBi
have two types of Fermi surfaces: one is a 2D hole Fermi surface
(Fig. \ref{fig1}(b)) processing such a shape of quite tetragonal
prism centered at the zone center $\Gamma$ and the other one is a 3D
electron Fermi surface (Fig. \ref{fig1}(b)) centered at the zone
corner $A$. These two 2D and 3D Fermi surfaces are obviously
originated from the No.2 and No.3 bands (see Fig. \ref{fig2}(b)),
respectively. In addition, we also noted that, from Fig.
\ref{fig1}(d) since the large gradient of the DOS around the Fermi
level which just locates at the valley, the superconducting
properties of NaBi may be highly affected by chemical impurities,
vacancies, and external strains.

\begin{table}
\caption{Superconducting parameters ($\lambda$-electron-phonon
coupling strength, $<\omega>_{ln}$-logarithmic average in cm$^{-1}$,
$T_c$-superconducting transition temperature in $K$, and
$\Theta_D$-Debye temperature in $K$) of NaBi without and with the SOC
inclusion. Note that Debye temperature has been derived according to
the elastic constants of NaBi with/without the SOC inclusion.}
\begin{ruledtabular}
\begin{tabular}{lcccccclllll}
    & $\lambda$ & $<\omega>_{ln}$ & $T_c$ & $\Theta_D$\\
\hline
non-SOC & 0.72 & 40.9 & 1.82-2.59 & 147.4\\
SOC     & 0.84 & 38.7 & 2.92-3.75 & 151.2\\
Expt\cite{Kushwaha}    & 0.62 &      & 2.15      & 140.0\\
\end{tabular}
\end{ruledtabular}
\label{tab1}
\end{table}

\begin{figure}[hbt]
\centering
\includegraphics[width=0.48\textwidth]{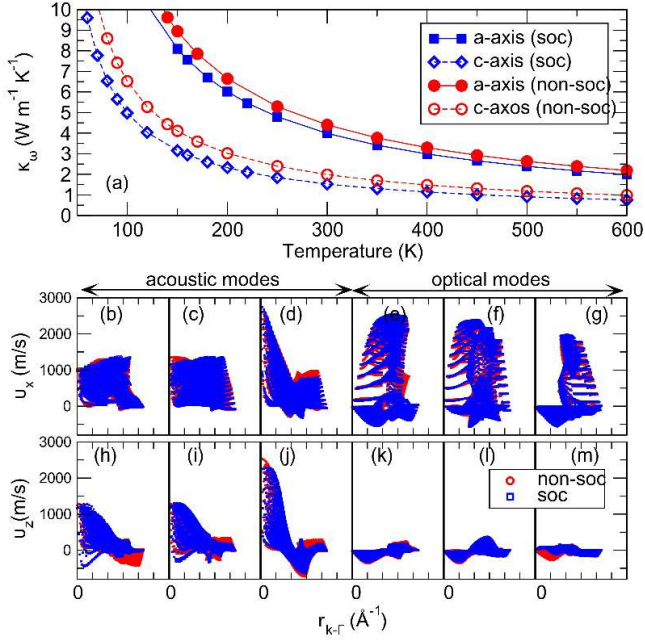}
\caption{
Upper panel: the derived lattice thermal conductivities along the
$a$- and $c$-axis with and without the SOC effect, respectively.
Lower panel: the phonon group velocities along the $a$ and $c$-axis
per phonon mode as a function of the distances between the given
\emph{k} point and the centered $\Gamma$ point (in total, 125 000
\emph{k}-point number in the BZ) with and without the SOC effect.}
\label{fig4}
\end{figure}

Furthermore, utilizing the frequency ($\omega$) of phonon and the
phonon group velocity ($v$) in the given transport direction and the
derived relaxation time ($\tau$) at wave vector $q$ and polarization
$j$ within the framework of linear Boltzman's equation, we further
derived the bulk lattice thermal conductivity as a function of
temperatures as follows,
\begin{equation}
\kappa_{\omega}
=\frac{1}{k_BT^2V}\sum_{q,j}n(q,j)[n(q,j)+1]\hbar^2\omega(q,j)^2v(q,j)^2_z\tau(q,j)_z
\end{equation}
where $k_B$, $V$, $n$ and $z$ are Boltzman constant, the crystal
volume, and Bose-Einstein distribution as well as the direction of
the thermal transportation. Specifically, the phonon's group
velocity and specific heat per mode have been derived according to
the second-order interatomic force constants obtained by the Phonopy
code \cite{Phonopy}. The relaxation time is in general determined by
third-order force constants, which are derivatives of the total
energy with respect to the atomic displacements in any three atoms
$i$, $j$, and $k$ in directions $a$, $b$, and $c$ within a large
supercell \cite{tau}. We have employed a real-space supercell
approach to anharmonic third-order force constant calculations using
the script of thirdorder.py \cite{wuli}, which analyzes the
symmetries of the crystal and significantly reduce the enormous
number of DFT runs that would be required to characterize all
relevant third-order derivatives of the energy. This method has been
successfully applied to calculate the lattice thermal conductivity
for a number of materials (such as, Si, diamond, InAs, and
lonsdaleite, \emph{etc}) \cite{diamond, si, wuli}.

Currently, our derived temperature-dependent lattice thermal
conductivities of NaBi have been compiled in Fig. \ref{fig4}. To our
surprising, it can been seen that NaBi exhibits apparent anisotropic
but extremely low lattice thermal conductivities. It has been also
noted that the SOC effect plays an important role in affecting the
$\kappa_\omega$. In comparison with the non-SOC case in Fig.
\ref{fig4}, the SOC effect heavily reduced the thermal
conductivities of both $a$- and $c$-axes directions. The mechanism
is mainly attributed to the that the SOC inclusion indeed results in
the softer phonon modes, as compared with those without the SOC
effect (Fig. \ref{fig3}b). At room temperature, along the $a$-axis
direction the $\kappa_\omega$ is found to be 4.40 Wm$^{-1}$K$^{-1}$
(3.98 Wm$^{-1}$K$^{-1}$), whereas along the $c$-axis direction the
$\kappa_\omega$ is extremely low, only about 1.98 Wm$^{-1}$K$^{-1}$
(1.53 Wm$^{-1}$K$^{-1}$) with (without) the SOC effect. In
particular, this low thermal conductivity is indeed comparable to
those of widely known materials with ultralow thermal conductivities
\cite{Tago,PbTe,PRX,SnSe}, such as PbS, PbSe, PbTe, PtLaSb, and
SnSe.

Interestingly, the low lattice thermal conductivity of NaBi exhibits
an obviously anisotropic ratio of $\kappa_\omega^{\rm
\emph{a}-axis}$/$\kappa_\omega^{\rm \emph{c}-axis}$ $\approx$ 2.2
(2.6) without (with) the SOC effect. Its anisotropy can be
interpreted well, according to the group velocities of the phonon as
illustrated in Fig. \ref{fig4}(b-m) in which the group velocities,
$v_x$ along the $a$-axis (Fig. \ref{fig4}(a-f)) and $v_z$ along the
$c$-axis (Fig. \ref{fig4}(b-m)), have been visualized as a function
of the \emph{k}-space distances between any phonon mode in the whole
BZ and the zone centered $\Gamma$ point. No matter whether the SOC
effect is included, both $v_x$ and $v_z$ show the quite similar
character, as evidenced in Fig.\ref{fig4}(b-m). The acoustic modes
play a main role in determining the lattice thermal conductivities.
In particular, along the \emph{a}-axis direction the phonon group
velocities are overall larger than those along the \emph{c}-axis
direction (Fig. \ref{fig4}(b-d) for a-axis and Fig. \ref{fig4}(h-j)
for c-axis), thereby resulting in a higher $\kappa_\omega$ along the
\emph{a}-axis. In general, the optical modes nearly makes no
contributions to the $\kappa_\omega$. However, based on our
calculations for NaBi, the optical modes make a certain contribution
to the $\kappa_\omega$ along the $a$-axis direction. For instance,
at room temperature along the $a$-axis direction the $\kappa_\omega$
that the optical modes contributed to  is about 0.822 (0.742)
Wm$^{-1}$K$^{-1}$ without (with) the SOC effect, being about 18\% of
the whole $\kappa_\omega$. This is mainly because its optical modes
exhibit very large group velocities along the \emph{a}-axis
direction (Fig. \ref{fig4}(e-g)) and, in the meanwhile, the low
frequencies (Fig. \ref{fig3}b) which are comparable to those of the
acoustic modes. In contrast, from Fig. \ref{fig4}(k-m) the optical
modes almost contribute nothing to the $\kappa_\omega$ along the
\emph{c}-axis direction. This fact further enlarges the anisotropic
ratio of the $\kappa_\omega$.

In summary, through first-principles calculations we have found that
NaBi is an intrinsic 3D TM with the combined properties of
electron-phonon induced superconducting and obviously anisotropic
but extremely low lattice thermal conductivity. The SOC effect has
been demonstrated to have the significant impacts on those
properties. Compared to topological insulators and topological
semimetals, our results for NaBi suggest that the topological metal
can be realized in a simple body-centred tetragonal structure
without any doping or strain treatments.

{\bf Acknowledgements} We thank Z. Fang for useful discussions. This
work was supported by the "Hundred Talents Project" of the Chinese
Academy of Sciences and from the Major Research Plan (Grand Number:
91226204) of the NSFC of China (Grand Numbers:51174188 and 51074151)
and Beijing Supercomputing Center of CAS (including its Shenyang
branch) as well as Vienna Supercomputing Center (VSC cluster).


\begin{thebibliography}{99}

\bibitem{Kane1} C. L. Kane and E. J. Mele,
Phys. Rev. Lett. \textbf{95}, 226801 (2005).

\bibitem{Shou-Cheng Zhang1} B. A. Bernevig, Taylor L.
Hughes, and Shou-Cheng Zhang, Science \textbf{314},1757 (2006).

\bibitem{Shou-Cheng Zhang3} X.-L. Qi, and S.-C. Zhang,
Rew. Mod. Phys.  \textbf{83}, 1057 (2011).

\bibitem{Kane2} M. Z. Hasan, and C. L. Kane,
Rew. Mod. Phys. \textbf{82}, 3945 (2010).



\bibitem{Dirac1} S. M. Young, S. Zaheer, J. C.Y. Teo, C. L. Kane,
E. J. Mele, and A. M. Rappe,
Phys. Rev. Lett. \textbf{108}, 140405 (2012).


\bibitem{Wang} Z. Wang, Y. Sun, X.-Q. Chen, C. Franchini, G. Xu,
H. Weng, X. Dai, Z. Fang, Phys. Rev. B \textbf{85}, 195320 (2012).

\bibitem{Cheng} X. Y. Cheng, R. H. Li, Y. Sun, X.-Q. Chen, D. Z. Li
and Y. Y. Li, Phys. Rev. B, \textbf{89}, 245201 (2014).

\bibitem{Wang2} Z. Wang, H. Weng, Q. Wu, X. Dai,
and Z. Fang, Phys. Rev. B, \textbf{88}, 205101 (2013).

\bibitem{Weyl1} H. Weyl, Z. Phys. \textbf{56}, 330 (1929).

\bibitem{Weyl2} H. B. Nielsen and M. Ninomiya,
Phys. Lett. B \textbf{130}, 389 (1983).

\bibitem{Weyl3} Y. Xu, R.-L. Chu, and C.W. Zhang
 Phys. Rev. Lett. \textbf{112}, 136402 (2014).

\bibitem{Weyl4} R. R. Biswas, and S. Ryu,
 Phys. Rev. B \textbf{89}, 014205 (2014).

\bibitem{Weyl5} T. Ojanen,
Phys. Rev. B \textbf{87}, 245112 (2013).

\bibitem{Weyl6} V. Pardo, J. C. Smith and W. E. Pickett,
Phys. Rev. B \textbf{85}, 214531 (2012).


\bibitem{Weyl8} D. Bulmash, C.-X. Liu, and X.-L. Qi,
Phys. Rev. B \textbf{89}, 081106(R) (2014).


\bibitem{Weyl10} X.G. Wan, Ari M. Turner, A. Vishwanath,
and S. Y. Savrasov, Phys. Rev. B  \textbf{83}, 205101 (2011).

\bibitem{Weyl11} T. Das, Phys. Rev. B  \textbf{88}, 035444 (2013).



\bibitem{Weyl14} J.-H. Zhou, H. Jiang, Q. Niu,
J.-R. Shi, Chin. Phys. Lett. \textbf{30}, 027101 (2013).

\bibitem{Murakami} S. Murakami,
New J. Phys. \textbf{9}, 356 (2007).

\bibitem{Young2} S. M. Young, S. Chowdhury,
E. J. Walter, E. J. Mele, C. L. Kane, and A. M. Rappe,
Phys. Rev. B \textbf{84}, 085106 (2011).

\bibitem{Liu} Z. K. Liu, {\em et al.}
Science \textbf{343}, 864 (2014).

\bibitem{Xu2013} S.-Y. Xu, {\em et al.}
Preprint at http://arXiv: 1312.7624 (2013).

\bibitem{Young} S. M. Young, S. Zaheer, J. C. Y. Teo,
C. L. Kane, E. J. Mele, and A. M. Rappe,
Phys. Rev. Lett. \textbf{108}, 140405 (2012).

\bibitem{Neupane} M. Neupane, {\em et al.}
Nature Communications, 5, 3786 (2014)

\bibitem{Borisenko} S. Borisenko, Q. Gibson, D.
Evtushinsky, V. Zabolotnyy, B. Buechner, R. J. Cava, Phys. Rev.
Lett., \textbf{113}, 027603 (2014).

\bibitem{ylchen2} Z. K. Liu, {\em et al.}
Nature Mater. \textbf{13}, 677 (2014)

\bibitem{Jeon} S. J. Seon, {\em et al.}, Nature Mater. \textbf{13},
(2014), DOI: 10.1038/NMAT4023

\bibitem{TM1-1D-Bi} J.W. Wells, J. H. Dil, F. Meier,
J. Lobo-Checa, V. N. Petrov, J. Osterwalder, M. M. Ugeda, I.
Fernandez-Torrente, J. I. Pascual, E. D. L. Rienks, M. F. Jensen,
and P. Hofmann, Phys. Rev. Lett. \textbf{102}, 096802 (2009).


\bibitem{TM3-half-Heusler} D. Xiao, Y.G. Yao, W.X. Feng,
J. Wen, W.G. Zhu, X.-Q. Chen, G. Malcolm Stocks, and Z.Y. Zhang,
Phys. Rev. Lett. \textbf{105}, 096404 (2010).

\bibitem{TM4-half-Heusler} H. Lin, L. AndrewWray, Y. Xia,
S.Y. Xu, S. Jia, R. J. Cava, A. Bansil and M. Z. Hasan, Nature
Mater. \textbf{9}, 546 (2010).



\bibitem{TM7-BiSb} D. Hsieh, D. Qian, L. Wray,
Y. Xia, Y. S. Hor, R. J. Cava, and M. Z. Hasan,
Nature (London),
\textbf{452},970 (4522008).


\bibitem{TM9-(Bi1-xInx)2Se3} M. Brahlek, N. Bansal,
N. Koirala, S.-Y. Xu, M. Neupane, C. Liu, M. Z. Hasan,
and O. Seongshik,
Phys. Rev. Lett. \textbf{109}, 186403 (2012).

\bibitem{TM10-TiS2-xTex} Z. Y. Zhu, Y.C. Cheng,
and U. Schwingenschlogl,
Phys. Rev. Lett. \textbf{110}, 077202 (2013).


\bibitem{Kushwaha} S. K. Kushwaha, J. W. Krizan,
J. Xiong, T.Klimczuk, Q. D. Gibson, T. Liang, N. P. Ong, and R. J.
Cava, J. Phys.: Condens. Matter \textbf{26} 212201 (2014).

\bibitem{Joseph} J. M. Reynolds, and C. T. Lane,
Phys. Rev. B, \textbf{79}, 405 (1950).

\bibitem{NaBi} E. Zintl and W. Dullenkopf,
Z. Phys. Chem. Abt. \textbf{16}, 183 (1932).

\bibitem{Fu2006} L. Fu and C. L. Kane,
Phys. Rev. B, \textbf{74}, 195312 (2006).

\bibitem{Fu2007} L. Fu and C. L. Kane,
Phys. Rev. B, \textbf{76}, 045302 (2007).

\bibitem{online}  See Supplemental Material
http://link.aps.org/supplemental/xxx. The supplemental materials
contains (1) the computational methods of electronic structure (VASP
\cite{vasp}), vibrational phonon dispersions (Phonopy \cite{Phonopy}
and QE \cite{qe}), electron-phonon coupling coefficients (QE
\cite{qe}), and lattice thermal conductivities (ShengBTE
\cite{wuli}) and (2) the detailed analysis of the band gap between
No.2 and No.3 bands in the whole BZ.

\bibitem{vasp} G. Kresse and J. Furthmu\"uller,
Phys. Rev. B \textbf{54}, 11169 (1996).

\bibitem{qe} P. Giannozzi, S. Baroni, N. Bonini, {\em et al.},
J. Phys.: Condens. Matter \textbf{21} 395502 (2009).

\bibitem{Phonopy} L. Chaput, A. Togo, I. Tanaka, and G. Hug,
Phys. Rev. B, \textbf{84}, 094302 (2011).

\bibitem{wuli} W. Li, J. Carrete, N. A. Katcho, N. Mingo,
Comp. Phys. Comm., \textbf{185}, 1747-1758 (2014).

\bibitem{MgB2} A. Y. Liu, I. I. Mazin,
and J. Kortus, Phys. Rev. Lett. \textbf{87},
087005 (2001).

\bibitem{ADF} P. B. Allen and
R. C. Dynes, Phys. Rev. B \textbf{12}, 905 (1975).

\bibitem{tau} A. Ward, D. A. Broido, D. A. Stwwart and G. Deinzer,
Phys. Rev. B, \textbf{80}, 125203 (2009).

\bibitem{diamond} L. Lindsay, D. A. Broido, and T. L. Reinecke,
Phys. Rev. Lett., \textbf{111}, 025901 (2013).

\bibitem{si} L. Chaput,
Phys. Rev. Lett., \textbf{110}, 265506 (2013).

\bibitem{Tago} J. M. Skelton, S. C. Parker, A. Togo, I. Tanaka, A.
Walsh, Phys. Rev. B, \textbf{89}, 205223 (2014).

\bibitem{PbTe} K. Biswas,
J.Q. He, I. D. Blum, C.-I. Wu, T. P. Hogan, D. N. Seidman, V. P.
Dravid, M. G. Kanatzidis, Nature, \textbf{489}, 414-418 (2012).

\bibitem{PRX} J. Carrete, W. Li, N. Mingo, S. D. Wang and S.
Curtarolo, Phys. Rev X, \textbf{4}, 011019 (2014).

\bibitem{SnSe}  L.-D. Zhao,
S.-H. Lo, Y.S. Zhang, H. Sun, G. J. Tan, C. Uher, C. Wolverton, V.
P. Dravid and M. G. Kanatzidis, Nature, \textbf{508}, 373-377
(2014).

\end{thebibliography}
\end{document}